%% file: RLJ.tex
\documentstyle[]{aipproc}

\input BoxedEPS
\SetRokickiEPSFSpecial  %% for dvips by Tom Rokicki, for VMS
\HideDisplacementBoxes
\input{mymak}
\renewcommand{\half}{{\textstyle\frac12}}
\newcommand{\fract}[2]{{\textstyle\frac{#1}{#2}}}

\begin{document}
\title{SPIN\\ Progress and Prospects\\
{\small\it A Talk Presented at SPIN2000,
October 30, 2000}}

\author{Robert~L.~Jaffe}
\address{Center for Theoretical Physics\thanks{This work is supported in part
by funds provided by the U.S. Department of Energy (D.O.E.) under cooperative
research agreement \#DF-FC02-94ER40818.\qquad MIT-CTP\#3075}\\
Massachusetts Institute of Technology\\
Cambridge, Massachusetts 02139}

%\lefthead{LEFT head}
%\righthead{RIGHT head}
\maketitle

\begin{abstract}
I review the progress in fundamental spin physics over the past several years
and the prospects for the future.  The progress is striking and the prospects
are excellent.
\end{abstract}

\pagestyle{plain}

\section*{Introduction}

I would like to thank the organizers for the honor of delivering the opening
talk at this Millennial Conference on Spin in High Energy and Nuclear
Physics.  Much is new, more will be forthcoming soon.  These
are exciting times.

Like this conference, my talk will focus on spin in QCD. The organizers
asked me to stress progress and prospects, which I will do. The prospects
for remarkable advances in the near future -- involving spin in one way or
another --  in electroweak unification and even in gravity compels me to
mention those fields as well.

\begin{figure}
$$
\BoxedEPSF{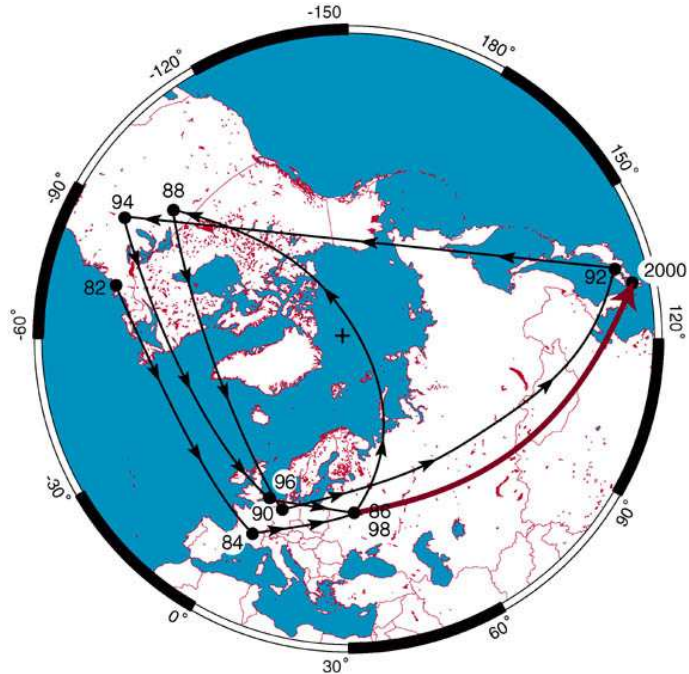 scaled 900}   
$$
\caption{The Prescott Effect.}
\label{RLJf1}
\end{figure}

In his welcome to SPIN98 in Protvino, Charlie Prescott, began his talk
by pointing out a striking geographical correlation between the
historical march of successive spin conferences and the Earth's angular
momentum~\cite{Prescott:1998}. Our location in Osaka once again
displays the ``Prescott Effect''.  I have updated the data in Fig.~\ref{RLJf1}.
 %
 %Figure 1 is the map of the earth you did for the conference  
%transparencies.
%
Word has it that the 2002 conference will move to Brookhaven, once again
confirming Prescott's Effect.  The rule, of course, displays the remarkably
international character of these conferences, which we all hope will
continue forever.  Parenthetically, as a theorist, I have to note the Anomaly
that occurred in 1986--1988, when the conference counter-rotated
from the USSR to the USA. Prescott appears to have fudged his data to
downplay this intriguing Anomaly -- it deserves further study.

Turning to more serious questions: The stuff of the world falls into
two categories: (1)~Gauge fields, which are bosons and are required by
local symmetries of space-time.  Gravitons follow from general
covariance.  $W$'s, $Z$'s, photons and gluons spring from local phase
invariances.  And (2)~Matter, which is composed of spin-$\onehalf$
particles, quarks, and leptons that carry the quantum numbers of
ungauged, global symmetries.  So far there are {\it no spinless
elementary particles at all!\/} One can argue that this pattern
follows in part from the constraints of renormalizability: Massive
vector bosons unrelated to gauge symmetries should not appear in our
low energy, renormalizable effective Lagrangian.  Spinless particles
might be propelled to Planck scale masses by quadratically divergent
self energies.  A couple of questions stand out: Why are there no
spinless elementary particles?  Why do only the fermions carry the
ungauged, global quantum numbers.  Why does matter exist at all, since
the (Yang-Mills and Einstein) gauge theories are quite consistent and
content without them?

Remarkably, we may be close to obtaining new experimental input
into these profound questions.
\begin{itemize}
	\item There is some evidence that a scalar Higgs boson awaits
	discovery at a mass around 115 GeV. Early indications at LEP will
	have to wait for FNAL or LHC for confirmation.  Of course
	radiative corrections to the Standard Model now constrain the
	Higgs mass reasonably well.  So unless Nature deals us a major
	surprise, the discovery of the first spinless elementary particle
	is imminent.
	\item A light Higgs suggests (but does not require) that
	supersymmetry is the natural extension of the Standard Model.  If
	so, we can expect to discover two entirely new forms of matter
	required by supersymmetry.  First, fermions without flavor
	associated with gauge symmetries -- the ``ino''s of SUSY like
	the gluino.  Supersymmetric partners of the gauge bosons, they
know nothing about the flavor symmetries of ordinary matter. 	
Second, the
scalar partners of quarks and leptons, which carry
	global flavor quantum numbers -- the ``sparticles'' of SUSY -- these
	would be the first bosons carrying the global symmetries which
	characterize matter.
\end{itemize}
 There are other good reasons to believe in supersymmetry:  partial
resolution to the hierarchy problem, coupling constant unification, and dark
matter candidates are most often mentioned.  It is worth remembering,
however, that SUSY introduces particles which are qualitatively different
from those we have known up to now.

Over the next few years, {\it spin\/} will play a central role in
testing and looking beyond the Standard Model and in exploring the
unresolved mysteries of QCD. Two examples will illustrate the
importance of spin in tests of the Standard Model:
	\bit \item The muon's magnetic moment, $(g-2)_{\mu}\equiv 2a_{\mu}$
	will be measured with nearly $20$ times existing precision by the
	Brookhaven $g-2$ experiment.   At this precision $a_{\mu}$ probes certain
	extensions of the Standard Model up to energies equivalent to
	 LEP and the Tevatron, and is sensitive to SUSY and other
	novelties.
	
	\item Electric dipole moments (EDM's) fascinate both theorists and
	experimenters.  They probe CP violation, one of the most poorly
	understood aspects of the Standard Model.  If all CP-violation is
	encoded in the CKM matrix, EDM's are too small to measure.  For
	this reason EDM's are an excellent place to look for CP-violation
	beyond the Standard Model.  The problem of baryogenesis in the
	early Universe continues to suggest that other sources of
	CP-violation are waiting to be discovered.
	\eit

Spin has recently proved itself a very powerful tool to probe the internal
structure of hadrons in QCD. We have measured the quark spin contribution
to the spin of the nucleon, but we do not understand it. In fact, we know
more about the spin of the {\it graviton\/}, which has never been
observed, than we do about the spin of the nucleon, which composes most
of the luminous mass in the Universe.  Several new results from the
Hermes collaboration at DESY and the SAMPLE collaboration at Bates whet
the appetite for future measurements of spin observables in QCD.  I will
review the outstanding issues in QCD spin physics in the latter 2/3 of this
talk.

Finally, I cannot fail to mention the experimental and technical foundations
on which our field rests.  Spin physics would go nowhere without the
extraordinary creativity and devotion of accelerators physicists, who have
developed novel methods of accelerating, storing and colliding polarized
particles and without experimentalists who have devised high density, high
polarization targets, and polarimeters and detectors capable of incredible
sensitivity.  Were not for this remarkable effort, we theorists might as well do
string theory.

\section*{Beyond the Standard Model}

 As an appetizer to QCD, which is the main course at this meeting, here is a
quick survey of some issues in spin physics beyond the Standard Model.

\subsection*{The spin of the graviton}

We know that the graviton has spin two.  Standard tests of general
relativity and the measured deceleration of the Hulse-Taylor pulsar assure
us of this.  Still, it would be nice to have direct observation of gravitational
radiation and explicit confirmation of its tensor nature.  LIGO, the Laser
Interferometry Gravitational Observatory, will do both if Nature is kind
enough to provide a strong enough source~\cite{Barish:1999vh}. LIGO I is
scheduled to begin data taking in 2003.  The upgrade to LIGO II, with much
greater sensitivity, begins in 2005.  Perhaps we shall see a direct
measurement of the spin of the graviton by the end of this decade.

\subsection*{The anomalous magnetic moment of the muon}

This subject is covered in the plenary talk by G.~Bunce, so I will be brief. 
After years of hard work and great patience, the Brookhaven experiment
(E821) seems poised to report a value for $a_{\mu}\equiv (g_{\mu}-2)/2$
which will challenge the Standard Model.  It is conventional to quote values
for $a_{\mu}$ in units of $10^{-10}$ or
$10^{-11}$ and accuracy in parts per million.  Thus the CERN $\mu^{+}$
value is $a_{\mu}\times 10^{10}=116\ 591\ 00(110)$ has an accuracy of 10
ppm~\cite{Bailey:1979mn}. The E821 number from 1998 running is
$a_{\mu}\times 10^{10}=116\ 591\ 91(59)$ (5 ppm), already a twofold
improvement in precision over the old CERN
experiment~\cite{Brown:2000sj}.

Theory includes electromagnetic, weak and strong corrections.  The pure
QED terms are known extremely well -- they contribute
$a_{\mu}$(QED)$\times 10^{11}=116\ 584\ 705.7(2)$.\footnote{See the talk
by G.~Bunce for the diagrams.} Strong (QCD) effects are very
significant: one-loop hadronic vacuum polarization gives
$a_{\mu}$(QCD-1~loop)$\times 10^{11}=6924(62)$; two-loop hadronic
vacuum polarization effects are important at the level of 1 ppm
($a_{\mu} (\mbox{QCD-2~loop})  \times 10^{11}=101(6)$);${}^2$ and QCD
light-by-light scattering enters at the level just below 1~ppm
($a_{\mu}\mbox{(QCD--light-by-light)}\times 10^{11}=-85(30)$).${}^2$  The
present precision of the theoretical estimate of
$(g_{\mu}-2)$ is principally limited by the lack of information on QCD
light-by-light scattering:
$a_{\mu}\mbox{(theory)}\times 10^{11}=116\ 591\ 62(8)$ (0.66 ppm).  Until
someone understands how to compute QCD light-by-light scattering more
accurately, there is no point carrying experiment beyond 0.5 ppm accuracy. 
This limit was designed into the BNL experiment: data on tape should allow a
precision of 0.5 ppm, and the experiment's ultimate goal is $\sim$0.35
ppm.  At this precision $a_{\mu}$ is sensitive to SUSY radiative corrections
from loops involving smuons and neutral and charginos, especially in
models with large
$\tan\beta$,
\begin{equation}
	a_{\mu}(\hbox{SUSY})\approx 140\times 10^{-11}
	\Bigl(\frac{100\hbox{GeV}}{\tilde m}\Bigr)^{2}\tan\beta 
	\label{eq1}
\end{equation}
so $a_{\mu}$ probes SUSY masses of order 100 GeV for $\tan\beta\sim 1$.
Of course, surprises beyond SUSY may await.

\subsection*{Electric Dipole Moments}

The search for an understanding of CP-violation probably commands more
resources than any other single issue in high energy physics:
$\varepsilon'/\varepsilon$, rare $K$ decays, $B$ factories, etc.  A classic
window into CP-violation is provided by the search for electric dipole
moments (EDM's).  Khriplovich's PANIC99 talk gives a good
summary~\cite{Khriplovich:2000zh}. I have abstracted Table~\ref{table1}
from his talk\footnote{Though the table does not do justice to the complexity
of measurements of nuclear EDM's or the sophistication of Khriplovich's talk}.
%
%Here belongs Table I. Taken from my talk.
\begin{table} 
\caption{Electric Dipole Moments}
\label{table1}
\begin{tabular}{lccc}
  Particle& Current Limit & Standard Model & Reasonable Goal\\
% \multicolumn{1}{c}{Previous result\tablenote{First demonstrate that
% state-of-the-model climate models are credible simulators of the true climate
% system.}} &
%   \multicolumn{1}{c}{Change\tablenote{Estimated 2.}(\%)}\\
\tableline
 Neutron & 6--$10\times10^{-26}$ & $10^{-30\mbox{--}31}$ &
$10^{-27\mbox{--}28}$\\[.5ex]
 Electron & $4\times10^{-27}$ & $10^{-40}$ &
$10^{-28}$\\[.5ex]
 Nuclei &$\sim 2\times10^{-24}$ & $\sim10^{-30}$ &
---\\
 Muon & $10^{-18}$ & $10^{-38}$ &
$10^{-24}{}^{\tablenote{Storage ring proposal, Y.~Semertzidis
\protect\cite{Semertzidis:1998sp}.}}$
\end{tabular}
\end{table}
Standard Model (ie.  CKM) predictions for EDM's are far smaller than the
reasonable goals of experiments. This means that EDM's provide fertile
ground in which to look for sources of CP-violation {\it beyond\/} the
Standard Model.  Of particular interest is Semertzidis's proposal to use the
BNL
$(g-2)$ ring to improve the limit on the muon EDM by as much as six orders
of magnitude~\cite{Semertzidis:1998sp}. Readers interested in this simple
and elegant idea should consult the review by Khriplovich.

\section*{Spin in QCD}

Polarization effects in QCD present a complex landscape.  Asymmetries
need to be explained. Sometimes we have no explanation but still can use
them to probe questions or isolate effects that are perhaps even more
interesting.  I want to highlight some of the topics I find particularly
interesting.  This week's program promises many interesting talks, beyond
my ability to anticipate -- my apologies to all those whose work has been
omitted in this brief overview.

I will focus on the overlap between theory and experiment.  Many
striking asymmetries occur in the low energy or nuclear domain where we
have few theoretical insights into QCD~\cite{Krisch:1998nw}. A few dramatic
spin-dependent effects occur in the deep inelastic domain, where QCD is
transparent.  Others occur where deep inelastic and soft domains overlap:
the world of parton distribution and fragmentation functions.  Here spin
effects help elucidate the puzzling nature of hadrons and here I will
concentrate.

The topics I will  cover include:
\begin{itemize}
	\item Bjorken's Sum Rule:  What it means to ``understand'' something 	in
QCD.
	\item Quark and gluon distributions in the nucleon.
	\item Probing polarized glue in the proton.
	\item The nucleon's total angular momentum:  Progress and 	frustration.
	\item Transversity.
	\item Spin at RHIC.
	\item Fragmentation and spin:  The new HERMES asymmetry and beyond.
	\item Spin dependent static moments:  $\mu_{s}$, the anapole, etc.
	\item The Drell-Hearn-Gerasimov-Hosada-Yamamoto Sum Rule
\end{itemize}

\subsection*{Bjorken's Sum Rule}

Occasionally it is worth reminding ourselves what it means to
``understand'' something in QCD. In the absence of fundamental
understanding we often invoke ``effective descriptions'' based on
symmetries and low energy expansions.  While they can be extremely
useful, we should not forget that a thorough understanding allows us to
relate phenomena at very different distance scales to one another. In
Bjorken's sum rule, the operator product expansion, renormalization group
invariance and isospin conservation combine to relate deep inelastic
scattering at high
$Q^{2}$ to the neutron's $\beta$-decay axial charge measured at very low
energy.  Even target mass and higher twist corrections are
relatively well understood.  The present state of the sum rule is
\bea
	\int_{0}^{1}dx\, g_{1}^{ep-en}(x,Q^{2})&=&
	\frac{1}{6}\frac{g_{A}}{g_{V}}\left\{1-\frac{\alpha_{s}(Q^{2})}{\pi}
	-\frac{43}{12}\frac{\alpha_{s}^{2}(Q^{2})}{\pi^{2}}
	-20.215\frac{\alpha_{s}^{3}(Q^{2})}{\pi^{3}}\right\}\nonumber\\
	&+&\frac{M^{2}}{Q^{2}}\int_{0}^{1}x^{2}dx\left\{
	\frac{2}{9}g_{1}^{ep-en}(x,Q^{2})+\frac{1}{6}
	g_{2}^{ep-en}(x,Q^{2})\right\}\nonumber\\
	&-&\frac{1}{Q^{2}}\frac{4}{27}{\cal F}^{u-d}(Q^{2})
	\label{eq2}
\eea
where the three lines correspond to QCD~\cite{Larin:1991tj}, target mass,
and higher twist\cite{Shuryak:1982pi} corrections respectively. $g_{1}$
and $g_{2}$ are the nucleon's longitudinal and transverse spin dependent
structure functions.  $g_{A}$ and $g_{V}$ are the neutron's
$\beta$-decay axial and vector charges.  ${\cal F}$ is a twist-4 operator
matrix element with dimensions of $[\hbox{mass}]^{2}$, which measures a
quark-gluon correlation within the nucleon,
\be
	{\cal F}^{u}(Q^{2})S^{\alpha}=\half\langle PS|\left.g\bar
u\widetilde{\mathbf 
F}^{\alpha\lambda}\gamma_{\lambda}u\right|_{Q^{2}}|PS\rangle
	\la{eq3}
\ee
where $g$ is the QCD coupling, ${\widetilde{\mathbf F}}$ is the dual gluon
field strength, and $|_{Q^{2}}$ denotes the operator renormalization point.

\begin{figure}
$$
\BoxedEPSF{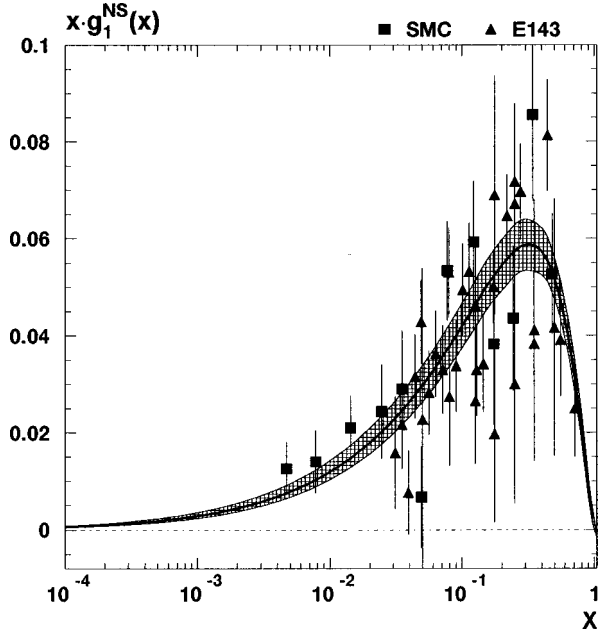 scaled 1000}   
$$
\caption{SMC analysis of data relevant to the Bjorken sum rule.}
\label{BjSR2}
\end{figure}

The most thorough analysis of the Bj Sum Rule I know of is one
presented by SMC in 1998~\cite{Adeva:1998vw}. Their theoretical
evaluation gives
\be
	\int_{0}^{1}dx\, g_{1}^{ep-en}(x,Q^{2})|_{\mathrm{theory}}=0.181\pm 0.003
	\la{Bjtheory}
\ee
at $Q^{2}=5$ GeV$^{2}$. Experiment is not yet able
to reach this level of accuracy.  The latest data relevant to the Bj
Sum Rule is shown in Fig.~\ref{BjSR2}.  The value extracted by the SMC
%
%This figure is Figure 10 from ref Adeva:1998vw.  See the figure 
% folder
is
\be
	\int_{0}^{1}dx\, g_{1}^{ep-en}(x,Q^{2})|_{\mathrm{expt.}}=0.174 \pm 0.005 
\begin{array}{ll}+0.011\\ -0.009\end{array}
	\begin{array}{ll}+0.021\\ -0.006\end{array}
	\la{Bjexpt}
\ee
at $Q^{2}=5$ GeV$^{2}$, and the errors are statistical, systematic,
and ``theoretical'' (eg.  generated by running the data to a common
$Q^{2}$), respectively~\cite{Adeva:1998vw}. Further accuracy is
necessary to confirm the target mass corrections and extract the twist-four
contribution.

\subsection*{Quark and gluon distributions in the nucleon}

No summary of recent progress in spin physics is complete without a
survey of the polarized quark and gluon distributions in the nucleon. 
These helicity weighted momentum distributions are the most precise and
interpretable information we have about the spin substructure of a
hadron.  The distributions are usually defined in terms of flavor-SU(3)
structure,
\bea
	\hbox{Singlet:}\qquad\Delta\Sigma &=& \Delta U +\Delta D +\Delta 	S\nonumber\\
	\hbox{Nonsinglet, isovector:}\qquad\Delta q_{3}&=& \Delta U -\Delta D \nonumber\\
	\hbox{Nonsinglet, hypercharge:}\qquad\Delta q_{8} &=& \Delta U +\Delta D -2\Delta 
S
	\la{qg-dist}
\eea
where $\Delta Q\equiv q^{\uparrow}(x,Q^{2})+\bar q^{\uparrow}(x,Q^{2})
-q^{\downarrow}(x,Q^{2})-\bar q^{\downarrow}(x,Q^{2})$.  Experimenters seem to prefer nonsinglet distributions specialized to the proton and neutron individually,
\bea
	\hbox{Proton nonsinglet:}\qquad\Delta q_{NS}(p)&=& \Delta U
-\half\Delta D
	-\half\Delta 	S\nonumber\\
	\hbox{Neutron nonsinglet:}\qquad\Delta q_{NS}(n)&=& \Delta D
-\half\Delta 	U-\half\Delta S 
	\la{qg-nonsinglet}
\eea
so that
\bea
	g_{1}^{p} &=& \fract{2}{9}\Delta\Sigma +\fract{2}{9}\Delta q_{NS}(p)
	\nonumber\\
	g_{1}^{n} &=& \fract{2}{9}\Delta\Sigma +\fract{2}{9}\Delta q_{NS}(n).
	\la{q-pn}
\eea

Since the integrated quark spin accounts for only about 30\% of the
nulceon's spin, it is extremely interesting to know whether the
integrated gluon spin in the nucleon is large.  Of course the
polarized gluon distribution, $\Delta g(x,Q^{2})$, cannot be measured
directly in deep inelastic scattering because gluons do not couple to
the electromagnetic current.  Instead $\Delta g$ is inferred from the
QCD evolution of the quark distributions.  [See
Ref.~\cite{Adeva:1998vw} for details of the process and references to
the original literature.]  However, evolution of imprecise data only
constrains a few low moments of $\Delta g$ and gives only crude
information on global characteristics such as the existence and number
of nodes.  It is clear that $\Delta g$ must be measured directly
elsewhere.

\begin{figure}[ht]
$$
\BoxedEPSF{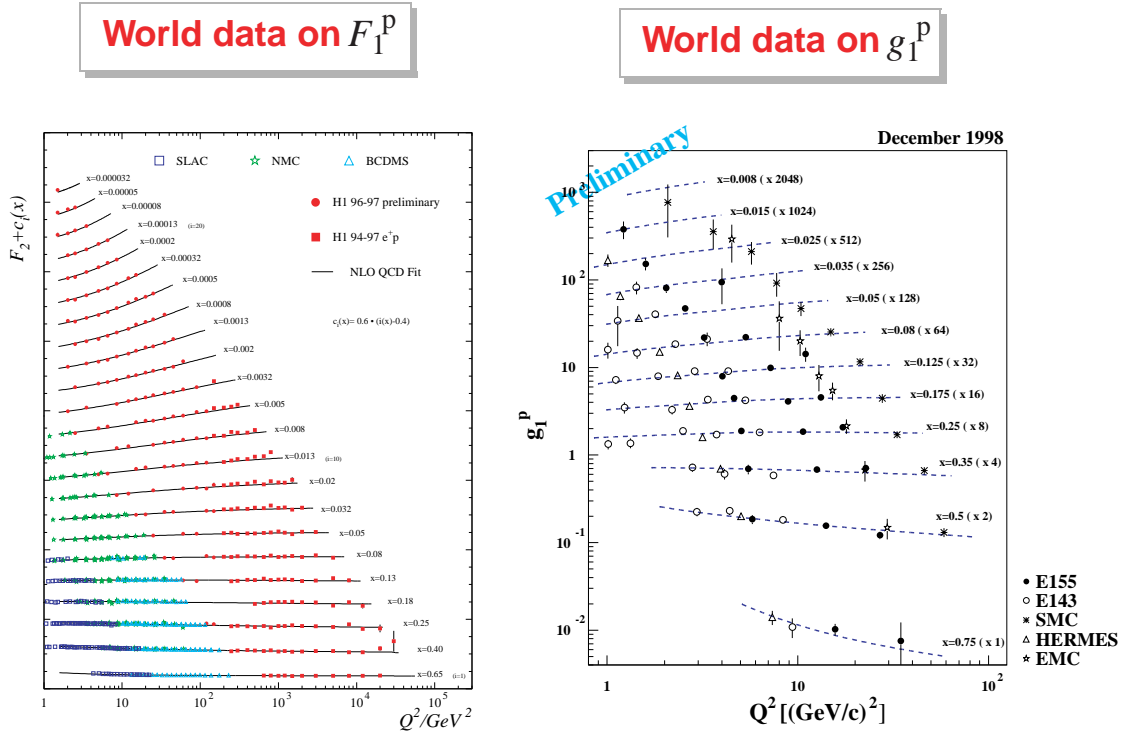 scaled 550}   
$$\smallskip
\caption{World data on spin-average and spin-dependent structure
functions~\protect\cite{Makins}.}
\label{Makins3}
\end{figure}

\begin{figure}[ht]
$$
\BoxedEPSF{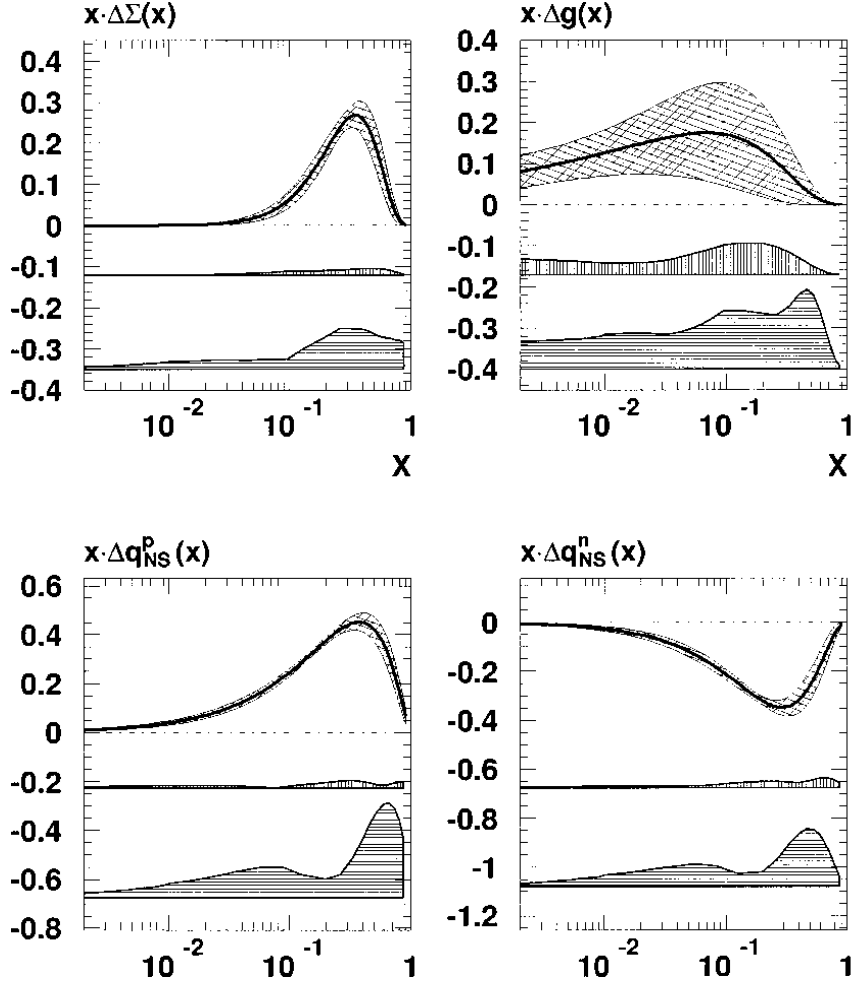 scaled 1600}  
$$\medskip
\caption{Polarized quark and gluon distribution functions. The upper figures
show the distribution with a statistical error bound. The lower figures show
estimates of systematic and theoretical uncertainties, respectively.}
\label{SMC4}
\end{figure}

That said, the world's data on polarized structure functions is summarized
in Figs.~\ref{Makins3} and
\ref{SMC4}.  Fig.~\ref{Makins3} is taken from Naomi Makins's talk at DIS2000
and presents the world's %
%This is a complicated two page figure taken from Naomi Makin's talk 
%at DIS2000.  It's called
%HermesDIS2000.pdf in your folder and it's 
%page 4 of the talk.  %
 data on $g_{1}^{p}$ in the same format traditionally used for unpolarized
structure function data~\cite{Makins}.  The figure highlights the 
tremendous progress of the past decade as well as the need for much better
data if our knowledge of polarized distributions would aspire to the same
accuracy as unpolarized distributions.  Fig.~\ref{SMC4} shows the quark and
gluon distributions extracted from the world's data by %
%This figure (SMC) is Fig. 5 from the same phys rev paper (Adeva et 
%al) mentioned before.
%
SMC, together with estimates of systematic and theoretical
uncertainties~\cite{Adeva:1998vw}. While the information on quark
distributions is fairly precise, it is clear that we know very little about the
distribution of polarized gluons in the nucleon.

\subsection*{Probing polarized glue in the nucleon}

As must have been clear from the preceeding discussion, a direct
measurement of the polarized gluon distribution in the nucleon is
probably the highest priority for groups interested in QCD spin
physics.  Further refinement of the indirect method championed by SMC
will contribute to this goal, but direct measurement is essential.

Several direct methods are being pursued:
\bit
	\item  $\bar c c$ pair production in $e\vec p_{\parallel}\to
	e' \bar c c X$ and related methods.
	
\begin{figure}[ht]
$$
\BoxedEPSF{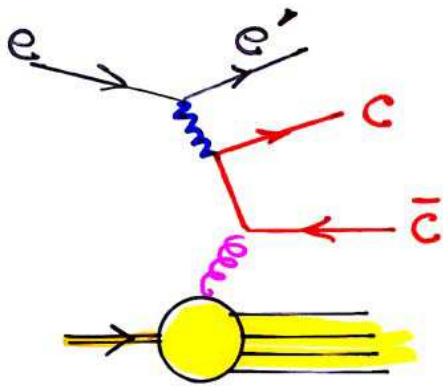 scaled 1000}  
$$
\caption{Measuring the polarized gluon distribution in $e\vec p_\parallel \to
e' c\bar c X$.}
\label{gluons5}
\end{figure}

	The COMPASS Collaboration has proposed to extend this powerful
	probe of the unpolarized gluon distribution to the polarized
	case~\cite{Bradamante:2000yc}. The basic mechanism is photon-gluon
	fusion, as shown in Fig.~\ref{gluons5}.  For further discussion see 	the talk
by Bradamante at this meeting.
	%
	%This is figure 2 from Bradamante's talk in your folder.

	Variations on this method include two jet production: $e\vec
	p_{\parallel}\to e' \hbox{ jet jet } X$ at large transverse momentum
	(as originally envisioned by Carlitz, Collins, and
	Mueller\cite{Carlitz:1988ab}); $\bar c c$ photoproduction $\gamma
	\vec p_{\parallel}\to \bar c c X$; and pion pair production,
	$\gamma\vec p_{\parallel}\to \pi\pi X$, which Hermes hopes to use
	a lower center of mass energies where $\bar c c$ and two jet
	production are not available~\cite{Vincter:2000pc}.
	
	\item Single photon production at high transverse momentum in
	polarized $\vec p_{\parallel} \vec p_{\parallel}\to \gamma \hbox{ jet } X$
 and
	related methods.
	
\begin{figure}[ht]
$$
\BoxedEPSF{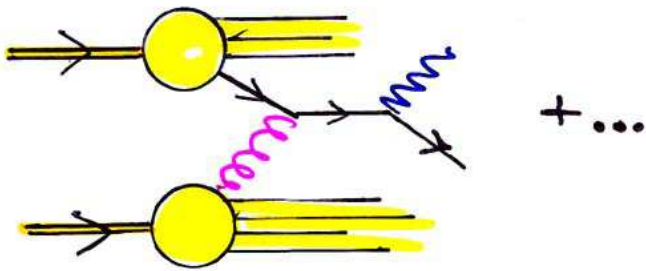 scaled 1000}  
$$
\caption{Measuring the polarized gluon distribution in $\vec p_{\parallel}
\vec p_{\parallel}\to\gamma\mbox{ jet } X$.}
\label{gluons6}
\end{figure}

	This is a prime goal for the polarized proton program at
	RHIC~\cite{Bunce:2000uv}. Here the basic mechanism is the QCD
	Compton process as shown in Fig.~\ref{gluons6}.  	%
	This process should be an excellent probe of the polarized gluon
	distribution.  However there is some controversy about higher-order QCD
corrections that has yet to be resolved in the
	unpolarized case. 
	Variations replace the high energy photon with a jet, or in the case 
of poor jet acceptance, a leading pion at high transverse momentum.
\eit
	%I'm sorry, but this figure has to be drawn in the manner of 	
%Bradamante's figure.  Perhaps you can just alter his.  I'll 	
%give you a drawing of what it must look like.
	%

\begin{figure}[hbt]
$$
\BoxedEPSF{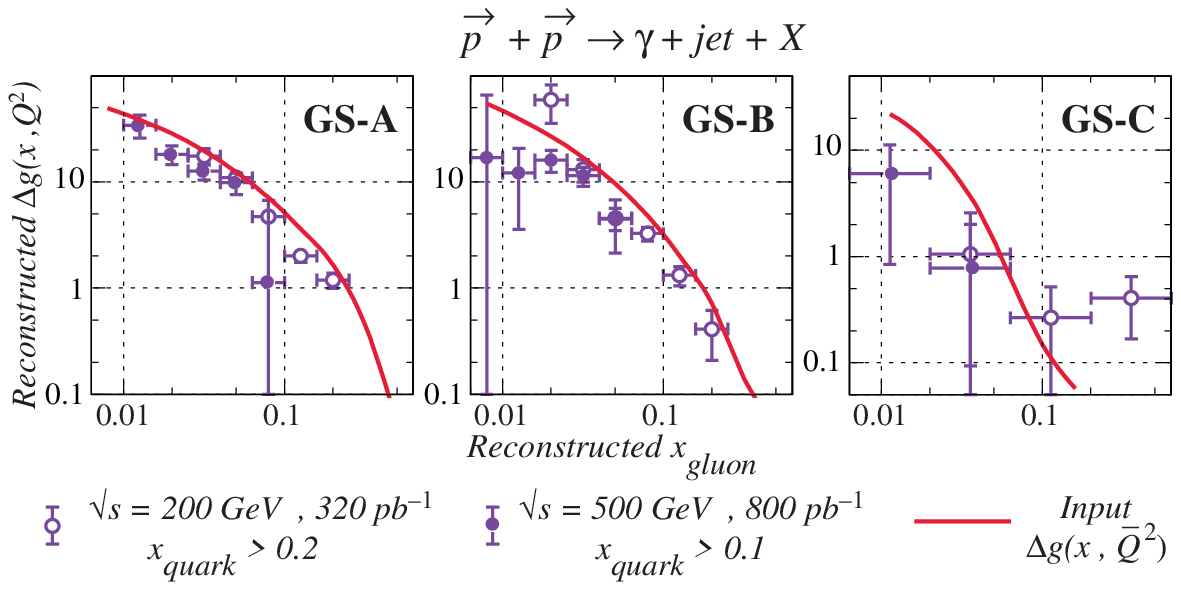 scaled 1200}  
$$
\caption{Estimates of polarized gluon distribution functions from 
$\vec p_{\parallel}
\vec p_{\parallel}\to\gamma\mbox{ jet } X$
at RHIC.}
\label{gluonexpt7}
\end{figure}

\begin{figure}[hbt]
$$
\BoxedEPSF{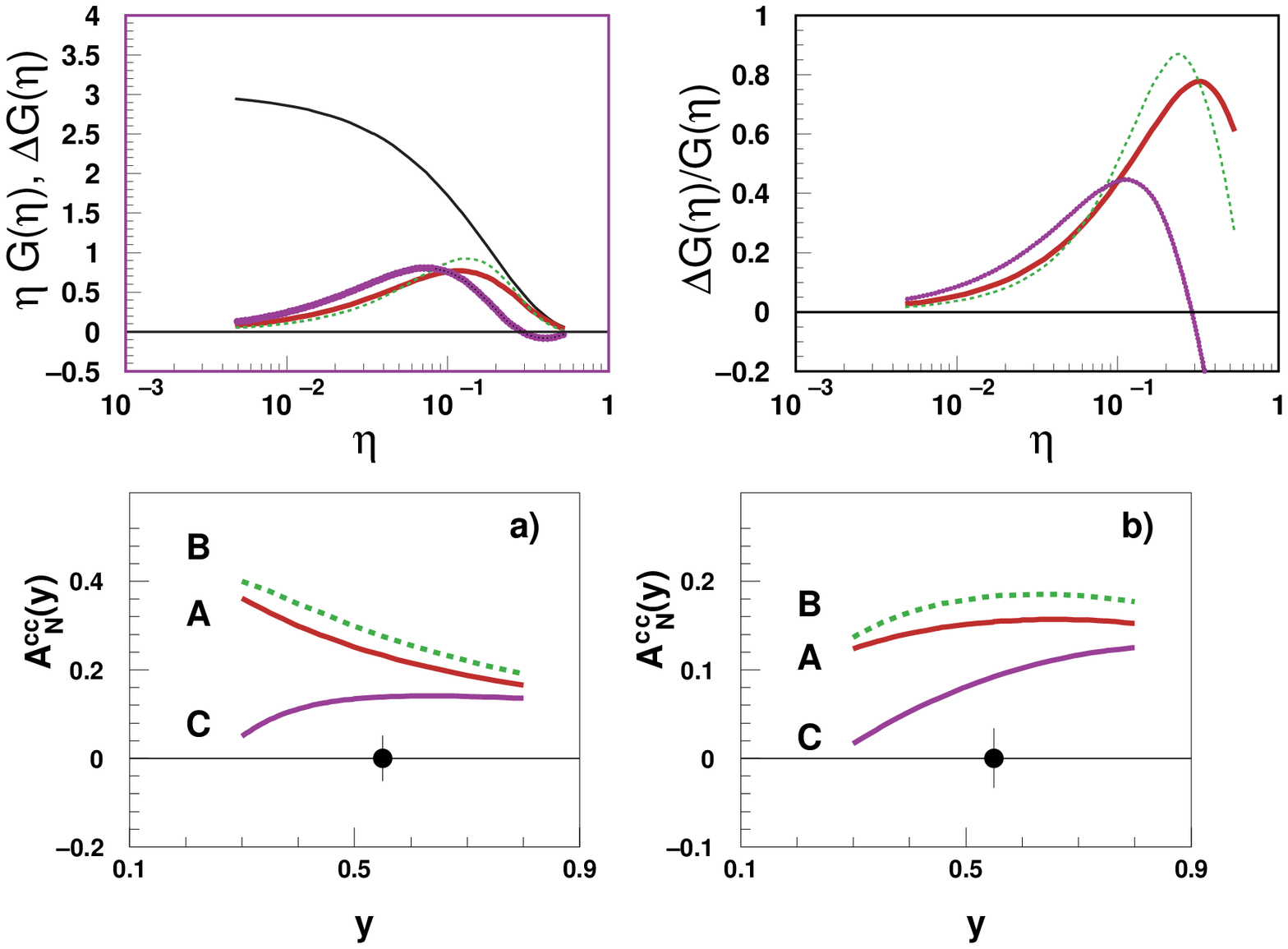 scaled 700}  
$$
\caption{Gluon-associated assymetries for $\gamma\vec p_\parallel \to
e' c\bar c X$ at COMPASS. See Bradamante's talk for details.}
\label{gluonexpt8}
\end{figure}

Estimates of the precision of these methods have become available as
better simulations come on line for COMPASS and RHIC. The projections
for $\vec p_{\parallel} \vec p_{\parallel}\to \gamma \mbox{ jet } X$ are
shown in Fig.~\ref{gluonexpt7}.
%
%This is Figure 8 from Ref.~Bunce:2000uv.  The pape is in your folder
%
An estimate of the COMPASS sensitivity is shown in Fig.~\ref{gluonexpt8}. 
%
%This is Figure 4 from Bradamante's article
%
In both cases the experiments are compared with gluon distributions
proposed by Gehrmann and Sterling.  Clearly, the polarized proton
program at RHIC has a major contribution to make in this area.

\subsection*{The nucleon's total angular momentum}

In the old days (pre-1988), it was clear that quark and gluon spin
distributions could be measured in deep inelastic scattering.  In some
uncertain sense they were imagined to be part of a relation that gave
the nucleon's helicity,
\be
	\half = \half\Delta\Sigma + \Delta g + \hbox{the rest}
	\la{oldspin}
\ee
where ``the rest'' was not well understood.  $\Delta\Sigma$ and $\Delta g$
were (and are) measurable, gauge invariant, and given by integrals over $x$,
$\Delta\Sigma=\Delta\Sigma(Q^{2})=\int_{0}^{1}dx \Delta\Sigma(x,Q^{2})$,
$\Delta g=\Delta g(Q^{2})=\int_{0}^{1}dx \Delta g(x,Q^{2})$.

Significant progress occured in the late  80s and 90s as the other
pieces of the angular momentum were related to local, gauge invariant
operators~\cite{Jaffe:1990jz}. This line of work culminated in Ji's
decomposition of the nucleon's helicity~\cite{Ji:1997ek},
\be
	 =\half\Delta\Sigma + \hat  L_{q} + \hat J_{g}
	\la{middlespin}
\ee
where $\hat L_{q}$ is the nucleon matrix element of an operator that
rotates quarks' orbital motion about the $\hat e_{3}$-axis in the rest
frame.  It is one candidate for a definition of the quark orbital
angular momentum in the nucleon.  $\hat J_{g}$ is the nucleon matrix
element of the operator that rotates the gluon about the $\hat e_{3}$
axis.  Ji showed that $\hat J_{g}$ cannot be further decomposed into
$\Delta g$ and an orbital contribution given by a local gauge
invariant operator.  This should not be too surprising because it is
well known that $\Delta g$ itself cannot be expressed in terms of a
{\it local\/} gauge invariant operator~\cite{Manohar:1991jx}. [In
general the operator is non-local, but becomes local in $A^{+}=0$
gauge.]  The virtue of eq.~\ref{middlespin} is that $\hat L_{q}$ can
be measured in deeply virtual Compton scattering (DVCS).  Although
$\hat J_{g}$ is in principle also measurable in DVCS, it requires a
precision study of $Q^{2}$ \emph{evolution} and is impossible in practice.

Most recently it has been possible to define gauge invariant {\it parton
distributions\/} for all the components of the nucleon's angular
momentum~\cite{Hagler:1998kg,Harindranath:1999ve,Bashinsky:1998if},
\be
	\half =\int_{0}^{1}dx\left\{ \half \Delta\Sigma(x,Q^{2})
   	+\Delta g(x,Q^{2}) + L_{q}(x,Q^{2}) +L_{g}(x,Q^{2})\right\}
	\la{newspin}
\ee
where $L_{q}$ and $L_{g}$ are Bjorken-$x$ distributions of quark and
gluon {\it orbital} angular momentum in the infinite momentum frame.
$L_{q}$ and $L_{g}$ are given by the light-cone fourier transforms of
bilocal operator products just like other parton distributions.  This
decomposition has many virtues: the four terms evolve into one another
with
$Q^{2}$~\cite{Hagler:1998kg,Harindranath:1999ve}, each term is the Noether
charge associated with the appropriate transformation of quarks or
gluons~\cite{Bashinsky:1998if}. Thus $L_{g}(x,Q^{2})$ is the observable
associated with the orbital rotation of gluons with momentum fraction $x$,
about the infinite momentum axis in an infinite momentum frame.  On the
other hand, eq.~\ref{newspin} suffers from a significant drawback: unlike
Ji's
$\hat L_{q}$, we know of no way to measure either $L_{q}(x,Q^{2})$ or
$L_{g}(x,Q^{2})$.  They do not appear in the description of DVCS.

So the situation with respect to a complete description of the nucleon's
angular momentum is frustrating.  The theory is under control. 
Eq.~\ref{newspin} summarizes all we would like to know, but we  do not
know how to measure what we would like to know. 

\subsection*{Transversity}

One of the major accomplishments of the recent renaissance in QCD spin
physics has been the rediscovery and exploration of the quark {\it
transversity distribution}.  First mentioned by Ralston and Soper in 1979
in their treatment of Drell-Yan $\mu$-pair production by transversely
polarized protons~\cite{RS1979}, the transversity was not recognized as a
major component in the description of the nucleon's spin until the early
1990's~\cite{artru90,jaffe91,cortes92,jaffe95}. We now know that the
transversity, $\delta q(x,Q^{2})$, together with the unpolarized distribution,
$q(x,Q^{2})$, and the helicity distribution,
$\Delta q(x,Q^{2})$, are required to give a complete description of the
quark spin in the nucleon at leading twist.  An equation tells this story
clearly --
\begin{eqnarray}
{\cal A}(x,Q^2) &=& \half q(x,Q^2)~I\otimes I + \half \Delta
q(x,Q^2)~\sigma_3 \otimes \sigma_3\nonumber\\[1ex]
&&\quad{} +\half  \delta q(x,Q^2)~
\left(\sigma_+\otimes\sigma_-+\sigma_-\otimes\sigma_+\right)
\label{symmetry}
\end{eqnarray}
Here, ${\cal A}$ is the quark distribution in a nucleon as a density matrix
in both the quark and nucleon helicities (hence the external product of two
Pauli matrices in each term).  $q$ governs spin average physics, $\Delta q$
governs helicity dependence, and $\delta q$ governs helicity flip -- or
transverse polarization -- physics.

\begin{figure}
$$
\BoxedEPSF{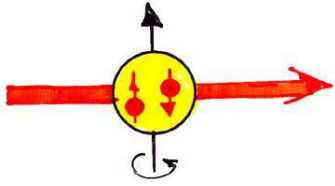 scaled 700}  
$$
\caption{Transversity: transversely polarized quarks in a nucleon at infinite
momentum.}
\label{transversity9}
\end{figure}

 The transversity can be interpreted in parton language as the probability
to find quarks of momentum fraction $x$, transversely polarized in a
transversely polarized nucleon at infinite momentum.  This is illustrated in
Fig.~\ref{transversity9}.
%
%This simple figure comes from my transparencies as marked

\begin{figure}[ht]
$$
\BoxedEPSF{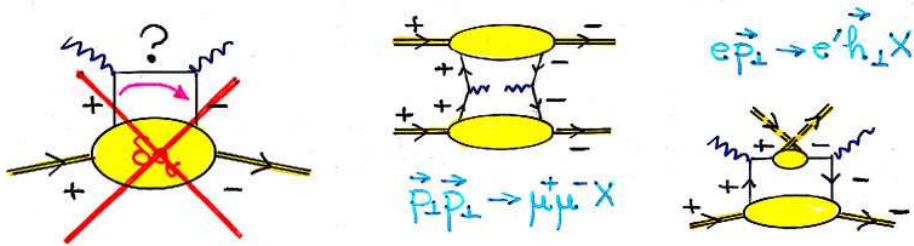 scaled 950}  %% 
$$
\caption{Deep inelastic processes relevant to transversity.}
\label{decouple10}
\end{figure}

The quark momentum distribution is well known and the helicity
distribution is becoming better known.  In contrast nothing is known
about transversity from experiment.  This is because it decouples from
inclusive DIS.  At leading twist
helicity and chirality are identical.  Transversity corresponds to
helicity (and therefore chirality) flip.  So transversity decouples
from processes with only vector or axial vector couplings.  This is
shown schematically in Fig.~\ref{decouple10}a.  In order to access
transversity it is necessary to flip a quark's helicity in one soft
process and compensate with another soft helicity flip process.  Two
examples where transversity does not decouple are transverse
Drell-Yan: $\vec p_{\perp} \vec p_{\perp}\to \mu^{+}\mu^{-} X$ (the
original Ralston-Soper process where transversity was discovered) and
semi-inclusive DIS where a final state fragmentation function flips
helicity: $e\vec p_{\perp}\to e'\vec h_{\perp}X$, of which several
examples exist.  Both are shown schematically in Fig.~\ref{decouple10}b and~c.

Measurements of quark transversity rank high on the agendas of Hermes,
COMPASS and RHIC.  Interest in this possibility has been piqued by the
azimuthal pion asymmetry recently reported by Hermes, as will be
discussed below.

\subsection*{Spin commissioning at RHIC}

One of the most important landmarks of the past year was the
successful spin commissioning run just concluded at RHIC. N.~Saito
will report in detail in his plenary talk.  Polarized protons were
accelerated in the AGS, and injected and stored in RHIC. The
Coulomb-nuclear interference polarimeter functioned as expected.  The
Siberian snake in RHIC rotated the polarization as planned.  Polarized
beam was accelerated in RHIC past depolarizing resonances and the
polarization was preserved with the aid of the snake.  
These milestones mark the beginnings of polarized collider physics at
RHIC, and a whole new window on the deep spin structure of hadrons.
This new facility would not have been possible without the support of
RIKEN and the effort of a team of Japanese experimenters working at
BNL and supported by the joint RIKEN/BNL Research Center.  Working
with both major RHIC detector groups, STAR and PHENIX, the RHIC Spin
Collaboration has developed an ambitious program for probing spin
structure in QCD.

\subsection*{Fragmentation and spin:\newline The Hermes asymmetry and
beyond}

To my mind the single most interesting development in QCD spin physics
reported since SPIN98 is the azimuthal asymmetry in pion
electroproduction from Hermes~\cite{Airapetian:2000tv}.  It is interesting
in itself and also as an emblem of a new class of spin measurements
involving spin-dependent fragmentation processes, which act as filters for
exotic parton distribution functions like transversity.

Fragmentation functions allow us to access and explore the spin structure
of unstable hadrons, which cannot be used as targets for deep inelastic
scattering.  Examples include the longitudinal and transverse spin
dependent fragmentation functions of the $\Lambda$, schematically $\vec
q_{\parallel}\to \vec \Lambda_{\parallel}$ and
$\vec q_{\perp}\to\vec \Lambda_{\perp}$.  Since the $\Lambda\to p \pi$
decay is self-analysing it is relatively easy to measure the spin of the
$\Lambda$.  By selecting $\Lambda$'s produced in the current
fragmentation region one can hope to isolate the fragmentation process
$q\to\Lambda$.  The principal challenge of such measurements is for
theorists: we have no theoretical framework for analysing fragmentation
functions.  Having measured the quark spin structure of the nucleon, we
can use flavor-SU(3) to estimate the way quark spins are distributed in the
$\Lambda$~\cite{Burkardt:1993zh}. However we do not know if this
information is reflected in the fragmentation process
$q\to\Lambda$.  Another, perhaps less obvious, example is the tensor
fragmentation function of the $\rho$, denoted schematically by
$(q\to\rho_{\pm}) - (q \to\rho_{0})$, where $\rho_{h}$ are $\rho$ helicity
states~\cite{Schafer:1999am}. $\rho$ decay transmits no spin information,
but it distinguishes the longitudinal and transverse helicity states required
for this measurement.  The data are already available.  The challenge to
theorists is to make use of it.

Even if we do not know how to interpret fragmentation functions, we can
use them as filters, to select parton distribution functions that decouple
from completely inclusive DIS.  The salient example is the use of a helicity
flip fragmentation function to select the quark transversity distribution.  As
shown in Fig.~\ref{decouple10}c, by interposing a helicity flip fragmentation
function on the struck quark line in DIS, it is possible to access the
transversity.  There are several candidates for the necessary helicity flip
fragmentation function:
	\bit

	\item $e \vec p_{\perp}\to e' \vec\Lambda_{\perp} X$
 	
	In this case the helicity flip fragmentation function of the
	$\Lambda$ is exactly analogous to the transversity distribution
	function in the nucleon~\cite{Kunne:1993nq,Jaffe:1996wp}. The only
	difficulty with this example is the relative rarity of $\Lambda$'s
	in the current fragmentation region, and the possibly weak
	correlation between the $\Lambda$ polarization and the
	polarization of the $u$ quarks, which dominate the proton.
	
	\item  $e \vec p_{\perp}\to e' \pi(\vec k_{\perp})X$\cite{Collins:1994kq}
	
	[The ``Collins Effect''] In this case the azimuthal angular
	distribution of the pion relative to the $\vec q$ axis can be
	analyzed to select the interference between pion orbital angular
	momentum zero and one that correlates with quark helicity flip. 	In
more traditional terms the effect is proportional $\vec
	S_{\perp}\cdot \vec q\times\vec p_{\pi}$.  This is multiplied by
	the quark transversity in the target and an unknown fragmentation
	function (known as the Collin's function) describing the
	propensity of the quark to fragment into a pion in a superposition
	of orbital angular momentum zero and one states.  The fact that
	fragmentation functions depend on $z$ while distribution functions
	depend on $x$ allows the shape of the transversity distribution to
	be measured in this manner.
	
	\item $e \vec p_{\perp}\to e' \pi\pi 	X$\cite{Collins:1994ax,Jaffe:1998hf}
	
	In this case the angular distribution of the two pion final state
	substitutes for the azimuthal asymmetry.
	
\eit

\begin{figure}
$$
\BoxedEPSF{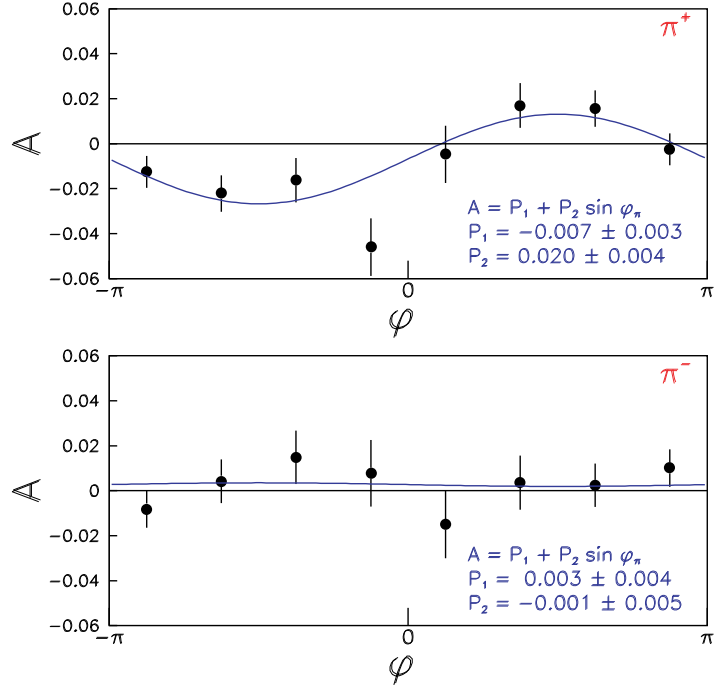 scaled 700}  
$$
\caption{The Hermes azimuthal asymmetry.}
\label{Hermes11}
\end{figure}

Last year Hermes announced the observation of an azimuthal asymmetry
similar to the Collin's asymmetry described above, but with a
longitudinally polarized target: $e \vec p_{\parallel}\to e' \pi(\vec
k_{\perp})X$.  Their data are shown in Fig.~\ref{Hermes11}. %
%This figure is the lower part (two graphs) from page 16 of Naomi 
%Makin's talk in HermesDIS2000.pdf in your folder
%
This asymmetry could be a (suppressed) reflection of the Collins
effect because the target spin, while parallel to the electron beam,
has a small component, ${\cal O}(1/Q)$ perpendicular to the virtual
photon.  It could also result from competing twist-three helicity flip
effects also suppressed by $1/Q$.  Unless the Hermes asymmetry is
entirely twist three, which seems unlikely, it appears that the
prospects for observing a large azimuthal asymmetry from a {\it
transversely\/} polarized target are very good.  Hermes will be
running with a transversely polarized target this year and their
results will be awaited with considerable excitement.

\subsection* {Spin-dependent static moments}

Polarization effects abound in the strong interactions at low energies.  Some
are quite striking, but most defy theoretical analysis because they occur in
two body scattering (or more complex processes), which we do not know how
to analyse in QCD.  One striking exception, quite similar in many ways to deep
inelastic physics, are the spin (and flavor) dependent static moments of the
nucleon.  In general these are measured in elastic lepton nucleon scattering,
$\ell N\to \ell' N$.  The general form is %
\be
	\Gamma \propto \langle PS| \bar q \Gamma q| PS\rangle
	\la{static}
\ee
where $\Gamma$ is some operator in the spin and/or space coordinates
of the quark field.  Familiar examples include the axial charges
($\Gamma =\gamma_{\mu}\gamma_{5}$), magnetic moments
($\Gamma=\half\vec r\times\vec\gamma$), and charge radii
($\Gamma=(\vec r)^{2}$).  A less familiar example is the ``tensor
charge'' ($\Gamma=\sigma^{0i}\gamma_{5}$), which though measurable in
principle, does not couple to any electroweak current and cannot be
measured in $\ell N\to \ell' N$.  The isovector, $u-d$, and
hypercharge, $u+d-2s$, flavor combinations are relatively easy to
measure given the variety of electroweak currents and baryons related
to one another by flavor-SU(3) transformations.  However, the third
flavor combination, $u+d+s$, does not appear in the electromagnetic or
charge-changing weak currents, and cannot be constructed by
flavor-SU(3) rotations because it is a flavor-SU(3) singlet and all
the other currents are flavor-SU(3) octets.

Much progress has been made in recent years both by theorists, who
have learned that the nucleon's tensor charge is related to the lowest
moment of its transversity structure function (in analogy to the
Bjorken Sum Rule); and by experimenters, who have measured the flavor
combination $u+d+s$ (and therefore the strangeness matrix elements) by
extracting the $Z^{0}$-nucleon coupling via parity violating $ep$
elastic scattering.  The $Z^{0}$ couples to weak isospin (hence
$u-d+s-c\ldots$) that, restricted to light quarks, is a linear
combination including the flavor singlet.
%
%Table Static is marked in the transparencies.
\begin{table} 
\caption{Electric Dipole Moments}
\label{table2}
\begin{tabular}{c|ccc}
\omit & \multicolumn{3}{c}{Dirac Structure}\\
\multicolumn{1}{c}{Flavor} & $\vec \gamma \gamma_5$ &
$\vec r \times \vec
\gamma$ &
$\sigma^{0i^{\mathstrut}}\gamma_5$ \\ \hline
$u-d$ & $\beta$-decay & Nucleon mag. mom. & Transverse DIS\\[1ex]
$u+d-2s$ & Hyperon $\beta$-decay & Hyperon mag. mom. & Transverse
DIS\\[1ex]
$u+d+s$ & Polarized DIS   & Parity odd $\vec c p\to ep$ & Transverse
DIS\\
\tableline
\omit & $q+\bar q$ & $q-\bar q$ & $q-\bar q$\\[1ex] 
\end{tabular}
\end{table}
Table~\ref{table2} shows a simplified summary of the Dirac and flavor
structure of some static matrix elements and how they are measured.

Two flagship measurements in this area are the extraction of
$\mu_{s}$, the nucleon matrix element of $s^\dagger \half\vec r\times\vec
\gamma s$, from parity violating $ep\to ep'$ (the SAMPLE experiment at
Bates), and extraction of $\langle r^{2}_{s}\rangle$, the nucleon matrix
element of
$s^{\dagger}(\vec r)^{2}s$, from the same process in a different
kinematic domain (the HAPPEX experiment at JLab).  Of course the
latter is not really spin-physics, but it belongs in the same
discussion.  HAPPEX first run at relatively large momentum transfer
saw no sign of a nucleon strange electric form
factor~\cite{Aniol:1999pn}. No statement can be made about
$\langle r^{2}_{s}\rangle$, however, until data at lower $Q^{2}$ become
available.

\begin{figure}[ht]
$$
\BoxedEPSF{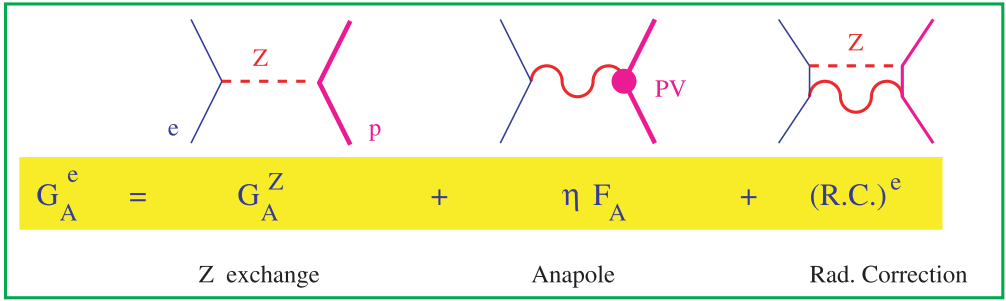 scaled 900}  
$$
\caption{Contributions to parity violation in $ep \to ep'$.}
\label{anapole12}
\end{figure}

\begin{figure}
$$
\BoxedEPSF{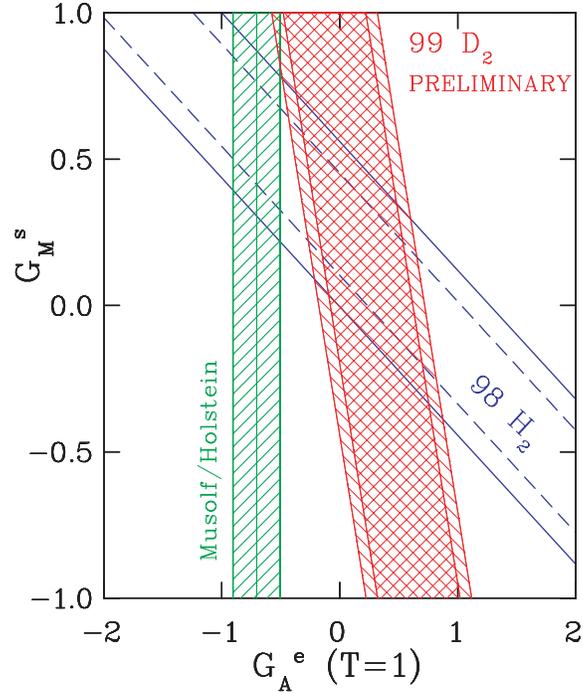 scaled 900}  
$$
\caption{Interpretation of SAMPLE measurements.}
\label{Sample13}
\end{figure}

SAMPLE's initial results are quite interesting -- for unexpected
reasons~\cite{Spayde:2000qg}.  The signal of interest, the $Z^{0}$-nucleon
coupling, is contaminated by a parity violating photon-nucleon interaction,
the so called ``anapole moment'', and by higher-order weak radiative
corrections to electron nucleon scattering (see Fig.~\ref{anapole12}). The
anapole and weak radiative corrections are not known a~priori and have to be
estimated in models~\cite{Musolf:1990ts,Musolf:1994tb}. They are
parameterized by $G_{A}^e (T=1)$ in Fig.~\ref{Sample13}.
A single measurement, say $\Pi\!\!\!\!/\ \ {ep\to ep}$, gives a line
in the $G_{M}^{s}\,$--$\,G_{A}^a (T=1)$ plane.  The initial SAMPLE
measurement together with the Holstein-Ramsey-Musolf estimate of
$G_{A}^a (T=1)$ gave a large positive estimate for $\mu_{s}(p)$
(admittedly with large error bars) in contrast to model calculations
that typically give negative $\mu_{s}$~\cite{Spayde:2000qg}. Most
recently, SAMPLE has announced measurements off a deuteron target,
$\Pi\!\!\!\!/\ \ {ed\to ed}$, which give an independent line in the
$G_{M}^{s}\,$--$\,G_{A}^a (T=1)$ plane~\cite{Mckeown}.  The result, shown in
Fig.~\ref{Sample13}b, %
%This is the data figure on the middle line of the PagesfromNSF4.pdf
%page.
suggests the estimate of $G_{A}^a (T=1)$ may be
wrong and that $\mu_{s}$ is closer to zero, though certainly
compatible with theoretical models predicting small negative values.

These are only the earliest results in what promises to be a productive study of the strangeness content of the nucleon using very precise measurements of parity violating elastic lepton nucleon scattering.
\subsection*{The Drell-Hearn-Gerasimov-Hosada-Yamamoto Sum
Rule}

The prospects for a definitive test of this deep and ancient sum
rule\cite{Gerasimov:1966et,Drell:1966jv,Hosoda:1966} are
now excellent.  Experiments proposed and/or underway at Mainz and JLab
will cover a wide range of energies with high polarization and high
statistics.  The question I would like to address here is ``What does
the DHGHY sum rule test?''.  The sum rule reads,
\be
	\frac{2\pi^{2}\alpha}{M ^{2}}\kappa^{2} = \int_{0}^{\infty} 	\frac{d\nu}{\nu}\left(\sigma_{P}(\nu)-\sigma_{A}(\nu)\right)
	\la{DHGHYI}
\ee
where $\kappa$ and $M$ are the anomalous magnetic moment and mass of
the target, and $\sigma_{P,A}$ are the total photoabsorption cross sections
(as functions of the laboratory photon energy, $\nu$) for target and photon
spins parallel and antiparallel.

The sum rule rests on two assumptions:
\bit
	\item Low's low-energy theorem
	
	Many years ago Low derived an extension of the Thompson limit in
	Compton scattering~\cite{Low:1954kd}. The nucleon's forward Compton
	amplitude can be written
	\be
		f(\nu)=f_{1}(\nu^{2})\vec\varepsilon'^{*}\cdot\vec\varepsilon
		+\nu f_{2}(\nu^{2})i\vec\sigma\cdot
		\vec\varepsilon'^{*}\times\vec\varepsilon
		\la{compton}
	\ee
	where $f_{1}$ and $f_{2}$ are the spin-nonflip and spin-flip	amplitudes
respectively.
	
	Using gauge invariance and QED, Low showed
	\be
		f_{2}(0)=-\half\frac{\alpha}{M^{2}}\kappa^{2}\ .
		\la{Low}
	\ee
	
	\item An unsubtracted dispersion relation
	
	Analyticity, crossing and unitarity dictate that the forward Compton 
amplitudes 	satisfies dispersion relations, which combine Cauchy's
theorem with 	the optical theorem (Im$f(\nu)\propto\sigma(\nu)$),
	\be
		\hbox{Re}f_{2}(\nu) =\sum_{j=0}^{J_{MAX}}c_{j}\nu^{2j}+
		\frac{1}{8\pi^{2}}\hbox{P}\int_{0}^{\infty}
		d\nu'^{2}\frac{\sigma_{A}(\nu')-\sigma_{P}(\nu')}
		{\nu'^{2}-\nu^{2}}
		\la{disp}
	\ee
	The polynomial is usually omitted in writing the dispersion
	relation, however it is not excluded by analyticity or unitarity. 	Since it
has no imaginary part, it does not affect the measurable
	cross sections.  For the moment let us omit the polynomial.  	
\eit
Then the DHGHY sum rule is obtained by evaluating $f_{2}(0)$ with the
aid of the dispersion relation, and equating it to Low's low energy
limit.

What could go wrong with this?  Absent any problems with
electrodynamics, the only weak point is ignoring the possible
polynomial in the dispersion relation.  Since we need only $f_{2}(0)$
only the constant term ($c_{0}$) in the polynomial matters.  Usually,
limits on the growth of amplitudes at high energies are invoked to
restrict the order of the polynomial.  However, they do not exclude
the constant, $c_{0}$.  In the standard derivations $c_{0}$ is simply
ignored.  This is called the assumption of an ``unsubtracted
dispersion relation''.  This is something of a misnomer: If the
integral eq.~\ref{disp} diverged it would be {\it necessary\/} to
reformulate it by formally subtracting $f_{2}(0)$ (remember we are
assuming $c_{1}=0, c_{2}=0,\ldots$).  The resulting integral would be
more convergent, but now the constant $f_{2}(0)$ would appear in the
relation,
\be
	\hbox{Re}f_{2}(\nu) =\hbox{Re}f_{2}(0)+
	\frac{\nu^{2}}{8\pi^{2}}\hbox{P}\int_{0}^{\infty}
	d\nu'^{2}\frac{\sigma_{A}(\nu')-\sigma_{P}(\nu')}
	{\nu'^{2}(\nu'^{2}-\nu^{2})}
	\la{disp''}
\ee
This is a ``subtracted dispersion relation''.  Now substitution into the Low's
theorem yields nothing useful.   Even if the dispersion relation does not {\it
need\/} subtraction, ie. even if the integral in eq.~\ref{disp} converges, still
the constant $c_{0}$ could be non-zero and spoil the sum rule.
Therefore, measurement of the high-energy behavior of $\sigma_A$ and
$\sigma_P$ \emph{does not} determine whether an additive constant  is
present in the dispersion relation. 

Debate about the validity of the DHGHY sum rule usually centers on
whether the dispersion integral needs subtraction.  Even if it doesn't, the
sum rule could be ruined by a non-zero, real constant in $f_{2}(\nu)$. 
Such a constant is called, for historical reasons, a ``$J=0$ fixed pole''.   So
the question of the validity of the DHGHY sum rule comes down to
whether $J=0$ fixed poles occur in QCD. It is known
that they do not occur in low orders of perturbation theory. This was first
verified when the electroweak anomalous magnetic moment of the muon
was first calculated using a generalization of these
methods~\cite{Altarelli:1972nc}.  It has been subsequently studied to higher
orders.  Brodsky and Primack have argued that it does not occur in ordinary
bound states\cite{Brodsky:1969ea} -- that the anomalous magnetic moment
of hydrogen can be calculated from a generalized DHGHY sum rule with out a
$J=0$ fixed pole.  Still, the verdict is out in QCD, where bound states are not so
simple.

If the DHGHY sum rule is verified experimentally this question will recede
to a footnote to history.  If, however, experiment fails to confirm it, we will
all have a lot to learn about $J=0$ fixed poles!

\section*{Conclusions}
My conclusions are brief.  We have made striking progress in recent years. 
The prospects for further progress are excellent.  I expect that spin physics
will continue to surprise us as it has in the past.  The reason is that spin is
fundamentally quantum mechanical in its origins so that it beggars our
classical intuition.  Remember -- we don't even know why
matter$\,\equiv\,$fermions exists!

\end{document}

%% file: mymak.tex
\newcommand{\be}{\begin{equation}}
\newcommand{\ee}{\end{equation}}
\newcommand{\bea}{\begin{eqnarray}}
\newcommand{\eea}{\end{eqnarray}}
\newcommand{\ben}{\begin{enumerate}}
\newcommand{\een}{\end{enumerate}}
\newcommand{\bit}{\begin{itemize}}
\newcommand{\eit}{\end{itemize}}
\newcommand{\la}[1]{\label{#1}}
\newcommand{\half}{\frac{1}{2}}

\def\pmb#1{\setbox0=\hbox{$#1$}%
\kern-.025em\copy0\kern-\wd0
\kern.05em\copy0\kern-\wd0
\kern-.025em\raise.0433em\box0 }